\documentstyle[emulateapj]{article}

\def\eps{\epsilon}
\def\epse{\eps_{e,-1}}
\def\epsB{\eps_{B,-2}}

\def\cm3{\mbox{cm}^{-3}}
\def\ergs{\mbox{ergs s}^{-1}}

\def\tdec2{t_{dec,2}}
\def\cm{\mbox{cm}}
\def\ergscm{\mbox{ergs s}^{-1}\mbox{cm}^{-2}}

\slugcomment{ }

\begin{document}

\title{GeV-TeV and X-ray flares from gamma-ray bursts}

\author{Xiang-Yu Wang\altaffilmark{1,2}, Zhuo Li\altaffilmark{3} and Peter M\'esz\'aros\altaffilmark{1,4}}
\altaffiltext{1}{Department of Astronomy and Astrophysics,
Pennsylvania State University, University Park, PA 16802;
xywang@astro.psu.edu, nnp@astro.psu.edu}
\altaffiltext{2}{Department of Astronomy, Nanjing University,
Nanjing 210093, China} \altaffiltext{3}{Department of Condensed
Matter Physics, Weizmann Institute of Science, Rehovot 76100,
Israel; lizhuo@wisemail.weizmann.ac.il} \altaffiltext{4}{Center
for Gravitational Wave Physics and Department of Physics,
Pennsylvania State University, University Park, PA 16802}
\begin{abstract}
The recent detection of delayed X-ray flares during the afterglow
phase of gamma-ray bursts (GRBs) suggests an inner-engine origin,
at radii inside the deceleration radius characterizing the
beginning of the forward shock afterglow emission. Given the
observed temporal overlapping  between the flares and afterglows,
there must be inverse Compton (IC) emission arising from such
flare photons scattered by forward shock afterglow electrons. We
find that this IC emission produces  GeV-TeV flares, which may be
detected by GLAST and ground-based TeV telescopes.  We speculate
that this kind of emission may already have been detected by EGRET
from a very strong burst---GRB940217. The enhanced cooling of the
forward shock electrons by the X-ray flare photons may suppress
the synchrotron emission of the afterglows during the flare
period. The detection of GeV-TeV flares combined with low energy
observations may help to constrain the poorly known magnetic field
in afterglow shocks. We also consider the self-IC emission in the
context of internal-shock and external-shock models for X-ray
flares. The emission above GeV from internal shocks is low, while
the external shock model can also produce GeV-TeV flares, but with
a different temporal behavior from that caused by IC scattering of
flare photons by afterglow electrons. This suggests a useful
approach for distinguishing whether X-ray flares originate from
late central engine activity or from external shocks.
\end{abstract}


\section{Introduction}

One of the key findings from the recently launched {\it Swift}
satellite is the common presence of X-ray flares in the early
afterglows of gamma-ray bursts (GRBs) (e.g. Burrows et al.
2005a,b; Zhang et al. 2005; Nousek et al. 2005; O'Brien et al.
2006). The flares typically occur at hundreds of seconds to hours
after the trigger, but in some cases days after the trigger. The
amplitude of the X-ray flare can be a factor of $\lesssim 500$ for
GRB050502B (Burrows et al.  2005; Falcone et al. 2005), and in
most cases a factor 3-10, compared with the background afterglow
component. Delayed GeV photons ($\sim 1.5 $ hours after the burst)
have been detected from GRB940217 by the EGRET, including one 18
GeV photon (Hurley et al. 1994). Motivated by the similarity
between this time delay in the GeV emission and that in the recent
X-ray flares, here we study the high energy signatures produced by
the inverse Compton (IC) scattering of the X-ray flare photons.

The rapid rise and decay behavior of some flares suggests that
they are caused by internal dissipation of energy due to late
central engine activity (e.g. Burrows et al. 2005a, Fan \& Wei
2005, Zhang et al. 2005; Wu et al. 2005). The observed overlapping
in time between flares and afterglows indicates that the forward
shock electrons which produce afterglows (M\'esz\'aros \& Rees
1997) are being exposed to these inner flare radiation, and must
produce IC emission by scattering the flare
photons{\footnote{Similar processes have been studied, such as the
IC scattering between the reverse shock photons (electrons) and
forward shock electrons (photons) (Wang et al. 2001), the IC
scattering of prompt MeV photons of GRBs by reverse shock
electrons  (Beloborodov 2005) and by afterglow electrons(Fan et
al. 2005).}}.  This process is expected to occur whether the X-ray
flares are produced by internal shocks or by other internal energy
dissipation mechanisms, as long as they occur inside the inner
edge of the forward shock region.

\section{IC scattering of X-ray flare photons by forward shock electrons}

We consider an X-ray flare of duration $\delta t$ superimposed
upon an underlying power law X-ray afterglow around time
$t=10^3t_3 \,{\rm s}$ after the burst, as observed in GRB050502B.
The X-ray afterglow is produced by a forward shock with  energy of
$E=10^{52}E_{52}{\rm erg}$ expanding in a uniform interstellar
medium with $n=1{\rm n_0 \,cm^{-3}}$. For GRB050502B, the prompt
burst fluence is $8\times 10^{-7} {\rm erg cm^{-2} }$. Assuming an
efficiency of $\sim10\%$ for the prompt gamma-ray emission, the
blast wave has an energy about $E=10^{52}{\rm erg}$ (for a burst
distance of $D=10^{28}\rm cm$.)  The X-ray flare occurs around
$t=10^3~ {\rm s}$ with a total fluence of $9\times 10^{-7} {\rm
erg~cm^{-2} }$ (Burrows et al. 2005a, Falcone et al. 2005).

From energy conservation of the adiabatic shock and $R=4
\Gamma^2ct$ (Waxman 1997), we can get  the Lorentz factor and
radius of the afterglow shock, i.e.
$\Gamma\simeq30(E_{52}/n_0)^{1/8}t_3^{-3/8}$ and $ R\simeq10^{17}(
E_{52} t_3/n_0)^{1/4}\cm$. From $R^2\Gamma^2U'_Xc=D^2F_X$, we have
the energy density of the X-ray flare photons in the forward shock
frame $U'_X=D^2F_X/(\Gamma^2R^2c)$, where $F_X$ is the observed
flare flux. This energy density is larger than the magnetic energy
density ($B^2/8\pi$) in the forward shock comoving frame when
$F_X>10^{-10} \epsilon_{B,-2}E_{52}t_3^{-1}D_{28}^{-2} \, {\rm erg
cm^{-2} s^{-1}}$, where $\epsilon_B$ is the equipartition factor
for  magnetic field in forward shocks. Since the flux of the X-ray
flare is also much larger than that of the underlying X-ray
afterglow { which may represent the synchrotron luminosity}, we
conclude that generally the afterglow electron cooling is
dominated by scattering  the X-ray flare photons.

The Lorentz factor of the electrons which cool by IC scattering of
X-ray flare photons in the dynamical time, $R/\Gamma c$, is$
\gamma_c={3m_e\Gamma c^2}/{4\sigma_TU'_X R}$, while the minimum
Lorentz factor of the post-shock electrons is
$\gamma_m\simeq1.5\times10^3{\epse
E_{52}^{1/8}}{n_0^{-1/8}t_3^{-3/8}}$ for a typical value of the
power-law index of the electron energy distribution $p=2.3$. Thus,
when the X-ray flare flux is larger than a critical flux
\begin{equation}
F_{X,c}=3\times10^{-10}E_{52}^{1/2}\epsilon_{e,-1}^{-1}n_0^{-1/2}t_3^{-1/2}D_{28}^{-2}\ergscm,
\end{equation}
we have $\gamma_m\ga\gamma_c$,  and all the newly shocked
electrons will cool, emitting most of their energy into  the IC
emission. This critical flux is usually much lower than the X-ray
flare flux averaged over its duration, so we conclude that the
flare photons can effectively cool the electrons in the forward
shock.

\subsection{IC emission and GeV-TeV flares}

As the forward shock propagates in the surrounding medium, the
energy that goes in to the newly shocked electron per unit
observer's time is $L_e=\epsilon_e L_{sh}=\epsilon_e 4\pi R^2 c U'
\Gamma^2$, where $U'=2\Gamma^2 n m_p c^2$ is the energy density of
the shock in the comoving frame. So, $L_e=2\times10^{48}\epse
E_{52}t_3^{-1}\ergs$. In the case where the characteristic flare
duration is comparable to the dynamic timescale $t$ of the
afterglow (i.e. $\delta t\sim t$) at the flare time, the received
IC luminosity is $L_{IC}\simeq L_e =2\times10^{48}\epse
E_{52}t_3^{-1}\ergs$ due to  fast-cooling of electrons,
corresponding to a flux at the detector of
\begin{equation}
F_{IC}=\frac{L_{IC}}{4\pi D^2 }\simeq10^{-9}\epse
E_{52}t_3^{-1}D_{28}^{-2}\ergscm
\end{equation}
But in the case where $\delta t< t=t_p$, where $t_p$ denotes the
peak time of the flare, the duration of the IC emission will be
lengthened by the angular timescale of the afterglow shock
($R/2\Gamma^2 c\sim t_p$) and therefore the received flux will be
reduced (though the total fluence does not change). The  total IC
energy is $E_{IC}=\delta t L_e=2\times
10^{51}\epsilon_{e,-1}E_{52}(\delta t/t_p) {\rm erg} $ and the
averaged IC flux is
\begin{equation}
F_{IC}\simeq10^{-9}\epse
E_{52}t_{p,3}^{-1}D_{28}^{-2}\left(\frac{\delta
t}{t_p}\right)\ergscm .
\end{equation}

 The incoming X-ray flare photons are likely to be anisotropic
seen by the isotropically distributed electrons in the forward
shock, so more head-on scatterings may decrease the IC emission in
the $1/\Gamma$ cone  along the direction of the photon beam
(corresponding to the angle less than $\pi/2$ relative to the
photon beam direction in the comoving frame), but enhance the
emission at larger angles with about half of the emission falling
into the angles between $1/\Gamma$ and $2/\Gamma$.  For a jet with
opening angle much larger than $1/\Gamma$, the jet geometry can be
approximately  regarded  as a sphere. Supposing that the flare
photons and external shock ejecta have a $4\pi$ solid angle (i.e.
not jet), then the IC emission in the observer frame should have
the same flux in every direction with a flux level as estimated
here.

 The angular dispersion of
high-energy IC photons will wash out the shorter temporal
structure of X-ray flares (Beloborodov 2005), so in both cases the
IC emission has a temporal structure determined by the forward
shock dynamic time, which could be quite different from the
temporal structure of the flare itself. The observed IC $\nu
F_\nu$ flux peaks at
\begin{equation}
\varepsilon_{IC,p}\simeq 2\gamma_m^2\varepsilon_X\simeq 3
\epse^2E_{52}^{1/4}n_0^{-1/4}t_3^{-3/4}\varepsilon_{X,\rm keV}
\mbox {GeV}
\end{equation}
where $\varepsilon_X$ is the  peak  of the  flare energy spectrum.
For the X-ray flare of GRB050502B, spectral fitting gives its
energy peak at $\sim \rm 2.5 keV$ (Falcone et a. 2005). The IC
emission occurs right inside the GLAST window and has a total
fluence about $10^{-7}$--$10^{-6}\epse E_{52}D_{28}^{-2}\rm erg \,
cm^{-2}$ for $0.1<\delta t/t<1$, so it should be detected by
GLAST. For the strongest bursts with $E\sim10^{54}\rm erg
\,D_{28}^2$, the GeV photons  could even have been detected by
EGRET, such as from GRB940217 (Hurley 1994).

If described approximately as broken power law, the IC energy
spectrum($\nu F_{\nu}$) has indices of $1/2$ and $-(p-2)/2$ before
and after the break at $\varepsilon_{IC,p}$ respectively. The
$-(p-2)/2$ power law spectrum can extend to a maximum energy
$\varepsilon_{IC, M}$, above which the IC falls into the
Klein-Nishina regime. Requiring the IC scattering to be in the
Thomson regime,$\gamma_e<\gamma_{e,M}= { {\Gamma m_e
c^2}/{\varepsilon_X}}$, gives
\begin{equation}
\varepsilon_{IC, M}=0.4
E_{52}^{1/4}n_0^{-1/4}t_3^{-3/4}\varepsilon_{X,\rm keV}^{-1}\,{\rm
TeV}.
\end{equation}
 The optical depth due to $\gamma\gamma$
absorption on the X-ray flare photons for the maximum energy
$\varepsilon_{IC, M}$ is $\tau_{\gamma\gamma}\simeq
0.3{F_{X,-9}n_0^{1/2}t_3^{1/2}}{E_{52}^{-1/2}}D_{28}^2({\delta
t}/{t})\varepsilon_{X,\rm keV}^{-1}$. Due to the low absorption
depth and the flat spectral slope above $\varepsilon_{IC, M}$, TeV
photons associated with bright X-ray flares are expected to be
detectable with  detectors such as H.E.S.S., MAGIC, VERITAS and
ARGO etc.

\subsection{Implications for the synchrotron afterglow }

The illumination by X-ray flare photons enhances the cooling of the
forward shock electrons, which in turn suppresses the synchrotron
afterglow emission. Thus, during the X-ray flare the optical afterglow
from the forward shock is expected to be much dimmer than the extrapolation
from the times before and after the flare.  The X-ray afterglow flux is
also suppressed during this period but this is masked by the X-ray flare.
Following Sari \& Esin (2000), we derive the synchrotron and
synchrotron self-Compton (SSC) luminosity when the external IC
seed photons (i.e. the X-ray flare photons here) are present.
 Denoting the ratio of the energy density of
the flare photons to that of the synchrotron photons by
$k=U_X/U_{syn}$, the luminosity ratios are
\begin{equation}
\frac{L_{SSC}}{L_{syn}}=\frac{L_{IC}}{k
L_{syn}}=\frac{U_{syn}}{U_B}=\frac{\eta
U_e}{U_B}(1+\frac{U_{SSC}}{U_{syn}}+\frac{U_{IC}}{U_{syn}})^{-1},
\end{equation}
where $\eta$ is the radiation efficiency of the electrons,
$L_{syn}$ ($U_{syn}$), $L_{SSC}$($U_{SSC}$) and $L_{IC}$($U_{IC}$)
are, respectively, the luminosities (energy densities) of
synchrotron, SSC and external IC emission. Defining the parameter
$Y\equiv({L_{IC}+L_{SSC}})/{L_{syn}}=(k+1)x_1$, we get $Y=\eta
(k+1){\epsilon_e}/{\epsilon_B}$ if $\eta
(k+1){\epsilon_e}/{\epsilon_B} \ll 1$ or $Y=[\eta
(k+1){\epsilon_e}/{\epsilon_B}]^{1/2}$ if $\eta
(k+1){\epsilon_e}/{\epsilon_B} \gg 1$. As $\eta\epsilon_e
L_{sh}=L_{syn}+L_{SSC}+L_{IC}$, we finally get
\begin{equation}
 L_{syn}=\frac{1}{Y+1}\eta \epsilon_e L_{sh},
  L_{IC}=k L_{SSC}=\frac{k}{k+1}Y L_{syn}
\end{equation}
Thus the synchrotron luminosity is reduced by a factor of $\sim
(k+1)^{1/2}$ (for  $\eta=1$), compared with the case without the
presence of a flare in this period. This effect should be
considered when modelling the part of the optical (or other
wavelength) afterglow that overlaps with X-ray flares (Gou et al.
2006). The SSC emission from the afterglow electrons is also
reduced, so GeV-TeV emission from this component  is lower than IC
emission of X-ray flare photons scattered by afterglow
electrons{\footnote{However, outside the period of the x-ray
flares, the SSC emission from the afterglow shock can produced a
background GeV emission, which could also be detectable by GLAST
under proper shock parameters(Zhang \& M\'esz\'aros 2001). }}.

In some bursts with X-ray flares, such as GRB050502B, the
underlying power law decay of the X-ray afterglow does not change
before and after the flare, which implies that the total energy
loss of the forward shock due to the IC emission is not
significant.  The total energy loss of the forward shock during
the flare period is $E_{\rm
loss}=\int_{t_1}^{t_2}L_{IC}dt=2\times10^{51}\epsilon_{e,-1}E_{52}{\rm
ln}(t_2/t_1) \, {\rm erg}$, where $t_1$ and $t_2$ are respectively
the beginning and ending times of the cooling period caused by the
flare, during which the flare flux is larger than $F_{X,c}$. Since
$E_{\rm loss}$ must be less than $E$, we get $\epsilon_e< 0.5/{\rm
ln}(t_2/t_1)$. For the case of the GRB050502B, the light curve of
its X-ray afterglow decays with a single power law  as
$t^{-0.8\pm0.2}$ before and after the flare. Taking
$F_{X,c}=0.3\times10^{-9}\epsilon_{e,-1}^{-1}{\rm {erg cm^{-2}
s^{-1}}}$ for this  burst, we can infer $\epsilon_e\la 0.3-0.4$
from the light curve of the flare.

{The microphysical shock parameter $\eps_B$ is currently less known in
the afterglow. Here we suggest that it can be constrained with multi-wavelength
observations from the optical, X-ray to the GeV-TeV range. With the observed IC
emission ($F_{IC}$), X-ray flare ($F_{X}$) and UV/optical afterglow ($F_{AG}$)
fluxes during the X-ray flare, we have the following constraint}
\begin{eqnarray}
\frac
{\eps_e}{\eps_B}\simeq\frac{U_{IC}}{U_B}=\frac{U_{IC}U_X}{U_XU_B}
=\frac{F_{IC}^2}{F_XF_{AG}}
\end{eqnarray}
where we have assumed $U_X>U_{syn}$ and $U_X>U_B$, and $F_{AG}$ is
the bolometric synchrotron flux of afterglow, which is emitted
mainly at
$\nu_m\simeq10^{15}E_{52}^{1/2}\epsB^{1/2}\epse^2t_3^{-3/2}$~Hz,
i.e., UV/optical band.

\section{Self IC scattering of X-ray flare photons  in late internal shocks}

{ Assuming that the X-ray flares are produced by late internal
shocks, we consider in this section the IC emission of the flare
photons by the same electrons that produce these photons, i.e. the
self IC emission.} In the late internal shock scenario, the exact
radiation mechanism producing the flares is unclear; it could be
synchrotron radiation or IC emission (Wei et al. 2006). If the
X-ray flares are produced by synchrotron emission, then the SSC
luminosity is
\begin{equation}
L_{SSC}=(\epsilon_e/\epsilon_B)^{1/2} L_{syn}
=(\epsilon_e/\epsilon_B)^{1/2} L_{\rm flare}
\end{equation}
where we have
assumed fast cooling for late internal shocks and
$\epsilon_e>\epsilon_B$. If instead the X-ray flares are produced
by first order IC emission, then the second order IC luminosity
(which is probably still in the Thomson regime) is
\begin{equation}
L_{IC, 2nd} =(\epsilon_e/\epsilon_B)^{1/3}
L_{IC,1st}=(\epsilon_e/\epsilon_B)^{1/3} L_{\rm flare}
\end{equation}
(Kobayashi et al. 2005). When the ratio $\epsilon_e/\epsilon_B$ is
not  too large, which is reasonable for internal shocks, the IC
luminosity is comparable to the flare luminosity in both cases.
Since internal shock may have a relative shock Lorentz factor of
the order of unity, i.e. $\Gamma_{\rm IS}\sim1$, the
characteristic Lorentz factor of electrons (in the comoving frame)
is $\gamma_{m, \rm IS}\simeq60 \epsilon_{e,-1}(\Gamma_{\rm IS}-1)$
for $p=2.5$, which is independent of the unknown bulk Lorentz
factor of the X-ray flares. So, the IC of the flare photons peaks
at
\begin{equation}
\varepsilon_{IC,p}=2\gamma_{m, \rm IS}^2
\varepsilon_X\simeq 10 \epsilon_{e,-1}^2(\Gamma_{\rm
IS}-1)^2\varepsilon_{X,\rm keV} \,{\rm MeV}.
\end{equation}
Except for the case when the late internal shock has a large
relative Lorentz factor $\Gamma_{\rm IS}$, this peak energy is
much lower than GeV, and therefore the contribution to the GeV-TeV
flux of this emission component is dominated by that of the IC
emission scattered by the forward shock electrons.

In the case when the X-ray flares come from the first-order IC
emission, we can even infer the corresponding synchrotron
luminosity, i.e.
 $L_{syn}\sim L_{\rm flare}$ when $\epsilon_e\ga\epsilon_B$. The
peak energy of this synchrotron emission is
$\varepsilon_{syn,p}=\varepsilon_X/(2\gamma_{m, \rm IS}^2)=0.1
\epsilon_{e,-1}^{-2}(\Gamma_{\rm IS}-1)^{-2}\varepsilon_{X,\rm
keV}\, {\rm eV}$. Thus infrared to optical observations during the
flare period give a potential approach to diagnose the emission
mechanism (i.e. whether synchrotron or IC) of  the X-ray flares.

\section{Self IC scattering of X-ray flare photons in the external shock scenario}

It has been suggested that  X-ray flares may come from an external
shock as it collides into a lumpy cloud in the surrounding medium
(Dermer 2005) or from the onset of the external shock (Piro et al.
2005; Galli \& Piro 2005). In this external shock scenario, the
flare photons come from the same region that produces the
afterglow, and the most likely emission mechanism is synchrotron
radiation. There is only a first-order IC process here, i.e. the
SSC process. Since the flare photons usually cause fast-cooling of
the electrons, this IC luminosity is
\begin{equation}
L_{SSC}= ({\epsilon_e}/{\epsilon_B})^{1/2}L_{\rm flare}.
\end{equation}
The peak energy of the SSC emission here is larger than that of
the SSC emission in the late internal shocks, because external
shocks have much larger shock Lorentz factors and hence much
larger characteristic electron energy. In the scenario where the
X-ray flares are produced by collision between the blast wave and
a lumpy cloud, the peak energy of the SSC emission is the same as
given by Eq. (4).  For the  scenario invoking the onset of
external shocks, the Lorentz factor $\Gamma$ is higher and the
peak likely locates around tens of GeV. Therefore in the external
shock scenario, the IC of X-ray flare photons can also produce
strong GeV flares that could be detected by GLAST.

The GeV flares produced in this case correlate tightly with the
X-ray flare, i.e., X-ray and high energy flares have similar
temporal profiles and durations, because they come from the same
emission region and electron population. On the other hand, in the
previous case of an afterglow IC emission due to inner-origin
X-ray flares illuminating the afterglow  electrons, the X-ray
flare and IC emission are produced in different regions, and the
duration of the high energy flare is determined by the outer
afterglow shock geometry, $\sim R/2\Gamma^2c\sim t$, leading to a
duration longer than that of the X-ray flare ($\delta t<t$). This
difference provides a useful approach to distinguish these two
different models of the X-ray flares, using future observations of
GeV-TeV flares.

\section{ Discussion and Conclusions}

Several models have been proposed to explain the delayed GeV
emission from GRB940217, including an electron IC emission
scenario in the forward shock (M\'esz\'aros \& Rees 1994, Zhang \&
M\'esz\'aros 2001), a hadron process scenario (e.g. Katz 1994;
Totani 1998) and a scenario invoking electromagnetic cascade
processes of TeV gamma-rays  in the infrared/microwave background
(Cheng \& Cheng 1996; Dai \& Lu 2002; Wang et al. 2004; Razzaque,
M\'esz\'aros and Zhang 2004). Here we have suggested a new model,
i.e. the electron IC emission accompanying the late time X-ray
flares recently discovered in GRBs. This model can be tested with
high energy detectors such as AGILE or GLAST operating
simultaneously with {\it Swift}. TeV gamma-rays accompanying
bright X-ray flares are also expected to be detected by
ground-based telescope such as H.E.S.S., MAGIC, VERITAS, ARGO. Due
to the delay times up to $10^3-10^4$ seconds of the X-ray flares
after the trigger, ground-based telescopes may have sufficient
time to slew to the directions of the bursts.

In summary, we have studied the IC emission associated with X-ray
flares in GRBs. We find the following results:\\
(1) { If the x-ray flares originate inside the forward shock
radius, e.g. from central engine activity, the IC emission due to
the X-ray flares overlapping with the forward shock electrons can
produce  GeV-TeV flares.} The fluence of such high energy flares
is sufficiently high for detection by high energy detectors such
as GLAST. The delayed GeV emission from GRB940217 could have been
produced in this process in a strong GRB. The light curves of such
GeV-TeV flares may  not correlate with that of the X-ray flares,
but will instead become
smoother and  longer.\\
(2) Strong X-ray flare fluxes can enhance the cooling of the
forward shock electrons, hence suppressing the synchrotron
afterglow emission during the periods of the flares. This will
cause dimming or changing of the decay slopes of the optical
afterglow.  Moreover, late-time GeV-TeV detection during X-ray
flares are useful for constraining the microphysics in the
afterglow shocks.\\
(3)GeV-TeV flares could also be produced by the self-synchrotron
Compton emission of the flare photons in an external shock
scenario of the X-ray flares. In this case, the light curves of
the GeV-TeV flares may correlate with the X-ray flares themselves.
So the different temporal behavior of the GeV-TeV flares for the
two IC processes { (1) and (3)} provides a potential way to
distinguish whether X-ray flares originate from a late central
engine activity or from an external shock scenario.

 \acknowledgments
We are grateful to E. Waxman, B. Zhang, Z. G. Dai,  S. Razzaque,
L. J. Gou, X. F. Wu and Y. Z. Fan for useful discussions, and the
referee for valuable comments. This work is supported by NASA
NAG5-13286 and NSF AST 0307376, an ISF grant and a Minerva grant
(for ZL), the National Natural Science Foundation of China under
grants 10403002, 10233010 and 10221001, and the Foundation for the
Authors of National Excellent Doctoral Dissertations of China (for
XYW).

\end{document}